\documentstyle[aps]{revtex}
\draft
\input epsf
\begin{document}
\twocolumn[\hsize\textwidth\columnwidth.85\hsize\csname@twocolumnfalse%
\endcsname
\title{Symmetric patterns of dislocations in Thomson's problem}
\author{A.\ P\'erez--Garrido$^1$ and M.\ A.\ Moore$^2$}
\address{$^{1\,}$ Departamento de F\'{\i}sica, Universidad de Murcia,
Murcia 30.071, Spain}
\address{$^{2\,}$ Theoretical Physics Group,
    Department of Physics and Astronomy,
   The University of Manchester, M13 9PL, UK}

\maketitle
\begin{abstract}
{
Determination of the classical ground state arrangement of $N$ charges 
on the
surface of a sphere (Thomson's problem) is a challenging numerical 
task. For
special values of $N$ we have obtained using the ring removal method of 
Toomre,
low energy states in Thomson's problem which have  icosahedral symmetry 
where
lines of dislocations run between the 12 disclinations which are 
induced by the
spherical geometry into the near triangular lattice which  forms on a 
local
scale.
 }
\end{abstract}
]
\narrowtext

Thomson's  problem consists of finding the ground state of $N$ Coulomb 
charges
confined to move on the surface of a sphere. While  this problem is 
simple to
specify its solution is not. It has been studied by many authors,
see \cite{EH91,AW94,PO96,MD96,EH97,PD97,AW97,PD97a} and references 
therein. On
a local scale, charges are disposed like those on a triangular lattice 
and
each charge has 6 nearest neighbors. On the other hand, Euler's theorem
guarantees the existence of at least 12 fivefold disclinations (charges 
with
only 5 nearest neighbors) on the sphere. More precisely, if $v_i$ is 
the number
of charges with $i$ nearest neighbors, then
\begin{equation}
\sum_i(6-i)v_i=12.
\label{euler}
\end{equation}

There exist methods to place the charges in  configurations with just 
12
disclinations each of which is at the corners of an icosahedron, see 
Ref.\
\cite{AW97}.  Those configurations are called icosadeltahedral and only 
exist
when  $N$ is given by
\begin{equation}
N=10(h^2+hk+k^2)+2,
\end{equation}
with $h$ and $k$ integers.

It was suggested in Ref.\ \cite{AW97} that these
configurations might be the ground states of Thomson's problem. Further 
work
showed, however, that configurations with dislocations (bound pairs of 
fivefold
and sevenfold disclination) have less energy than icosadeltahedral
configurations \cite{PD97,PD97a}, as lines of dislocations emanating 
from the
disclinations  act to screen the disclinations by reducing their 
strains fields
\cite{DM97}. However, we were unable to find any  patterns of 
icosahedral
symmetry containing
dislocations \cite{PD97}. It is the purpose of this Brief Report to 
show that
such patterns can be found if one uses the ``ring removal" technique of 
Toomre
\cite{TOOMRE}.

\begin{figure}
\epsfxsize=.85\hsize
\begin{center}
\leavevmode
\epsfbox{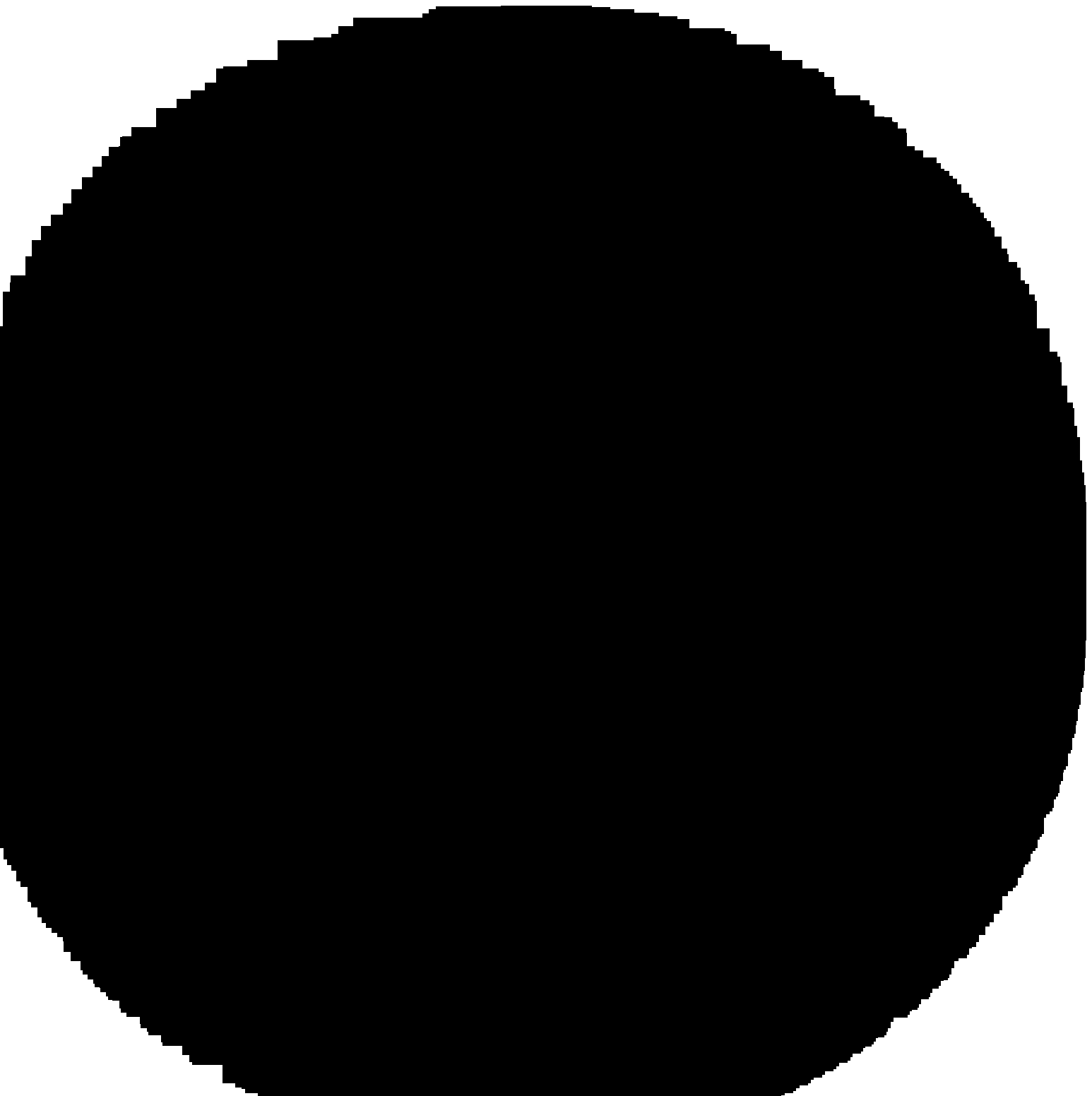}
\end{center}
\caption{Initial configuration with the 3rd and 7th rings removed for 
an
initial icosadeltahedral with 12002 charges ($h=k=20$).}
\label{fig1}
\end{figure}

Each disclination in an icosadeltahedral configuration is surrounded by 
rings
of $5n$ charges, where $n$ is the order number of the ring. By removing 
the
charges in one of these  rings and then  relaxing the energy of the
system it is possible to obtain a configuration with 5 dislocations
symmetrically disposed around the disclination. Removing several rings 
we can
get lines of dislocations which act to screen the disclination as 
predicted by
Dodgson and Moore \cite{DM97}. One must be careful how one  chooses  
the rings
to be removed; if one removes consecutive rings the final 
configurations are
not usually symmetric. In Fig.\ \ref{fig1} we have plotted an 
icosadeltahedral
configuration with $h=k=20$, i.e. 12002 charges, after removing the 3rd 
and 7th
rings around each disclination. This type of initial configuration was 
used in
conjunction with  standard numerical procedures (mostly the conjugate 
gradient
method) to minimize the interaction energy $E$ of the Coulomb charges 
with each
other.

\begin{figure}
\epsfxsize=0.85\hsize
\begin{center}
\leavevmode
\epsfbox{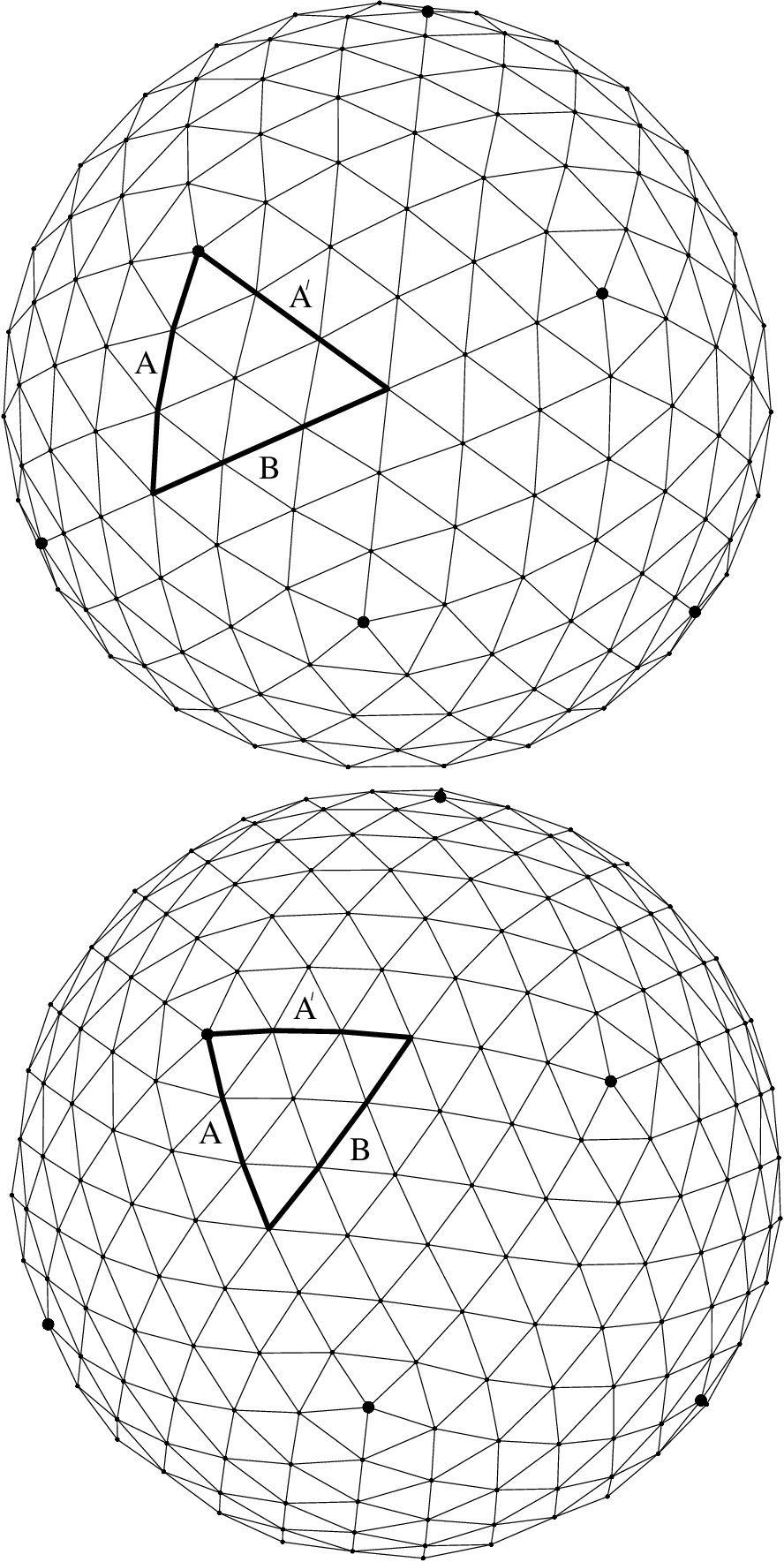}
\end{center}
\caption{Icosadeltahedral configurations for low number of particles.
Disclinations are represented by solids dots. Each line links two 
nearest
neighbors particles. Thicker lines  indicate a region (facet) which 
would be
equilateral but for the spherical geometry, for the cases $h=k=3$ 
(figure at the top) and $h=6$, $k=0$ (figure at the bottom).}
\label{figab}
\end{figure}

It is possible to  estimate the total number of rings $n_r$ to be 
removed
around each disclination (but not the actual ring numbers themselves,
unfortunately) as follows. In Fig \ref{figab}  are shown  examples of 
regions
(which we shall refer to as facets)
for icosadeltahedral configurations which naively one would expect to 
have
three equal sides but which cannot achieve this because of the 
spherical
geometry.

Let us denote by $A$ and
$A^\prime$  the length of the equal sides of the facets and $B$ the 
remaining
one. $A=A^\prime<B$. These lengths can be calculated by simple 
geometric
arguments. For the case $h=k$ one has
\begin{equation}
A=R\tan^{-1} \left(\frac{\alpha/2}{\cos \frac{2\pi}{10}}\right),
\end{equation}
and
\begin{equation}
B=R\cos^{-1} \left( 1-0.690983\sin^2A/R\right),
\end{equation}
where $R$ is the radius of the sphere and  $\alpha$ is the angle 
between two
neighboring
disclinations where
\begin{equation}
\alpha=2\tan^{-1}\left(\frac{\sqrt{5}-1}{2}\right).
\end{equation}
For $h>k=0$, $A$ is given by
\begin{equation}
A=\frac{R\alpha}{2}.
\end{equation}
One should note that the above is only exact for $h$ even. When $h$ is 
odd, $A$
is slightly larger but since the difference tends to zero as $h$ 
increases we
shall not take it into account.

To triangulate each facet one subdivides $A$ and $B$ into $D_A$ and 
$D_B$
segments respectively. In Fig \ref{figab} $D_A$ and $D_B$ are both 
equal to
$3$, but this need not be the case for larger values of $N$. To find 
the number
of rings to take out, $n_r$, one requires that $A/D_A$ be as close as 
possible
to $B/D_B$, since then the spacing between charges will  be most 
uniform,
thereby minimizing the strain field energy caused by the spherical 
geometry.
The difference between
$D_A$ and $D_B$ is then the number of rings to remove, $n_r$. With this 
in mind
we  obtain the following expression for $n_r$:
\begin{equation}
n_r={\rm Round}\left[ \left(1-\frac{A}{B}\right)D_B \right],
\end{equation}
where the function Round[$x$] gives the closest integer  to $x$. $D_B$ 
is
related to the number of particles $N$ by the relation
\begin{equation}
D_B=\sqrt{\frac{N-2}{p}},
\end{equation}
for the case $h=k$, $p=30$ while for $h>k=0$, $p=40$. Thus, within this
approximation, one is able to estimate out how many rings to remove for 
given
$h$ and $k$.  In
Tables 1 and 2 we show for both cases the value of $N$ at which $n_r$ 
changes
its value. These estimates of $N$ are consistent with our observations
 on systems with up to 16000 particles. We have not studied the
general case $h>k>0$ as it is difficult to identify the facets to 
triangulate.

\begin{table}
\caption{This table shows the smallest value of $h$ at which  the 
number of
rings to be removed should be increased by one and $n_r$, the total 
number of
rings to be removed for
particle numbers $N=30h^2+2$, ($h=k$) in the original icosadeltahedral 
state.}
\begin{tabular}{lll}
$h=k$ & $N$  &$n_r$\\
\tableline
           5 &          752 &            1 \\
          15 &         6752 &            2 \\
          24 &        17282 &            3 \\
          34 &        34682 &            4 \\
          43 &        55472 &            5 \\
          52 &        81122 &            6 \\
          62 &       115322 &            7 \\
          71 &       151232 &            8 \\
\end{tabular}
\label{tabla1}
\end{table}

\begin{table}
\caption{This table shows the smallest value of $h$ at which  the 
number of
rings to be removed should be increased by one and $n_r$, the total 
number of
rings to be removed, for  particle numbers
$N=10h^2+2$, ($k=0$) in the original icosadeltahedral state.}
\begin{tabular}{lll}
$h$, $(k=0)$ & $N$  &$n_r$\\
\tableline
          5 &         1002 &            1 \\
          13 &         6762 &            2 \\
          22 &        19362 &            3 \\
          30 &        36002 &            4 \\
          38 &        57762 &            5 \\
          47 &        88362 &            6 \\
          55 &       121002 &            7 \\
          64 &       163842 &            8 \\
\end{tabular}
\label{tabla2}
\end{table}

We next describe how we are going to report our values for this energy. 
Using
Ewald sums one can  calculate the energy for charges on an infinite 
plane
triangular lattice \cite{BM77} and deduce that for the sphere with unit 
radius
and for unit charges \cite{EH91}
\begin{equation}
2E=N^2-1.1061033... N^{3/2}+...,
\label{energy2}
\end{equation}
as $N\rightarrow\infty$.  It is useful therefore to study $E_i$ given 
by
\begin{equation}
E_i=\frac{2E-N^2}{N^{3/2}},
\end{equation}
which will approach  $-1.1061033...$  as $N\rightarrow\infty$.
We will refer to $E_i$  as the ``energy" of the system.

\begin{figure}
\epsfxsize=.85\hsize
\begin{center}
\leavevmode
\epsfbox{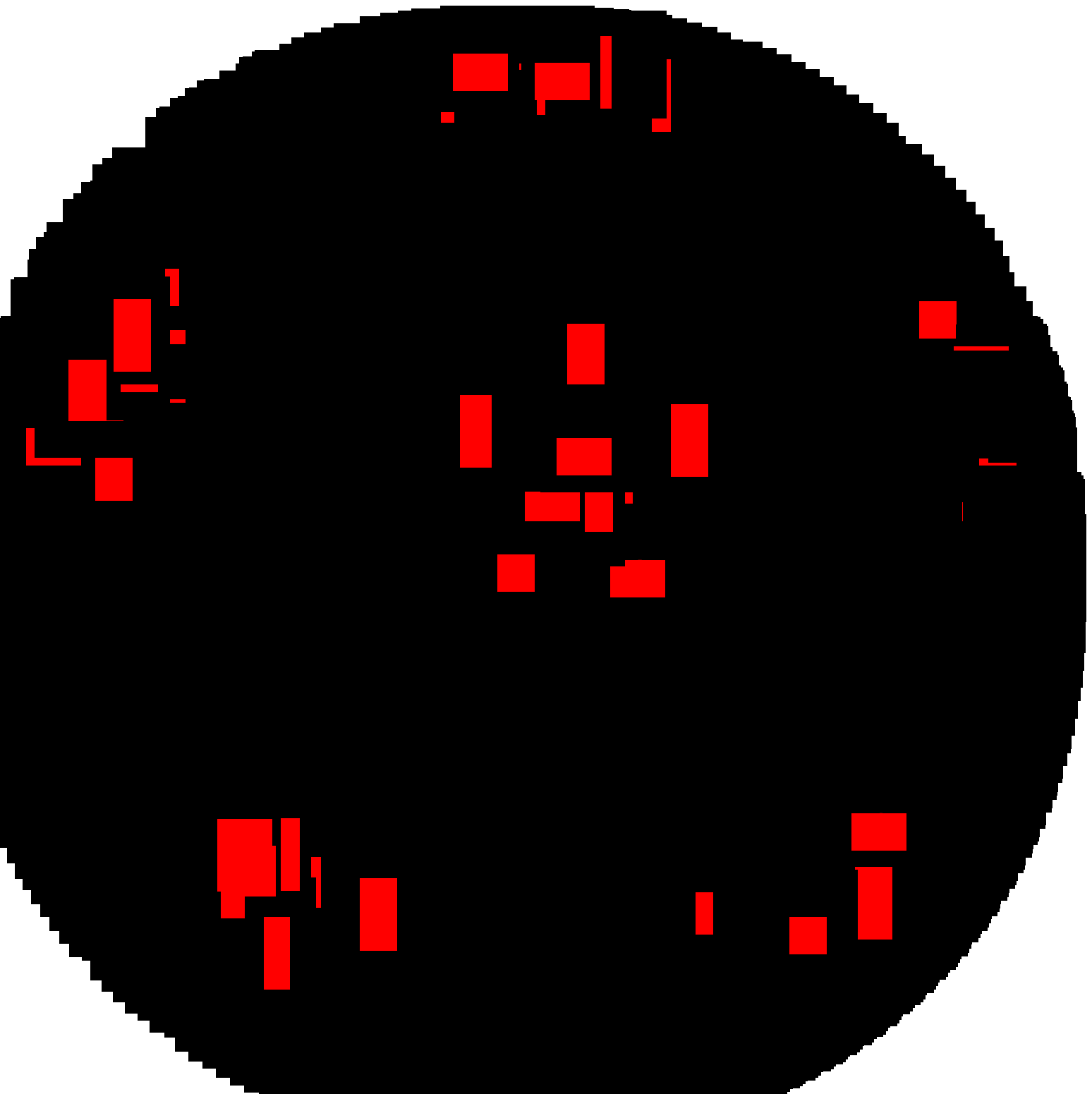}
\end{center}
\caption{11342 charges after relaxation. Largest spots correspond to 
seven-fold
disclinations, medium spots represent five-fold disclinations and 
smallest
spots are normal six-fold coordinated charges. This configuration is 
obtained
after removing the 2nd and 9th rings ($\equiv 1320$ charges)  from the
icosadeltahedral configuration with 12002 charges ($h=k=20$). The 
energy of
this configurations is \hbox{$E_i=-1.10558942$.} }
\label{fig2}
\end{figure}

\begin{figure}
\epsfxsize=.85\hsize
\begin{center}
\leavevmode
\epsfbox{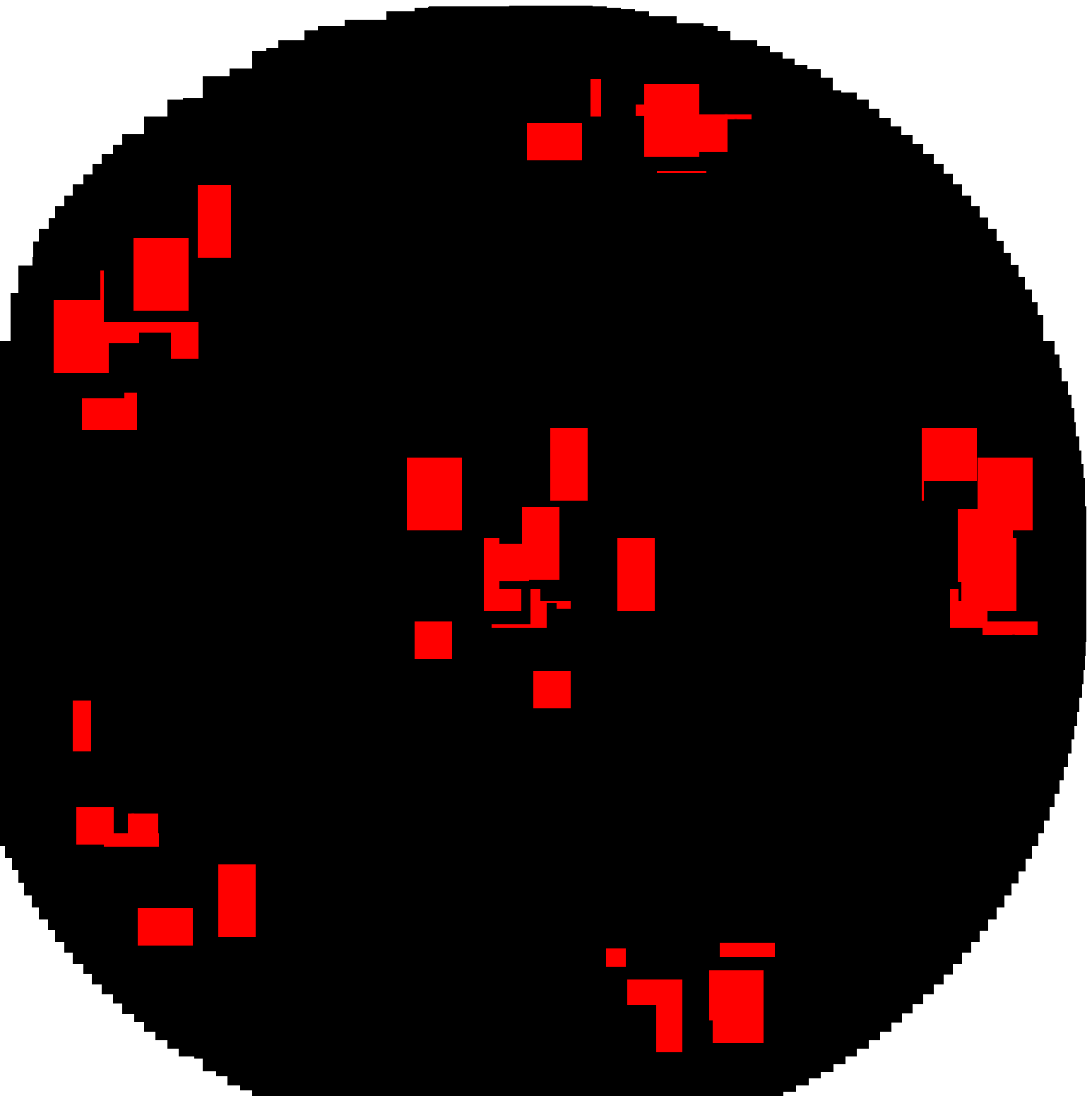}
\end{center}
\caption{15152 charges after relaxation. Largest spots correspond to 
seven-fold
disclinations, medium spots represent five-fold disclinations and 
smallest
spots are normal six-fold coordinated charges. This configuration is 
obtained
after removing the 2nd and 10th rings ($\equiv 720$ charges)  from  the
icosadeltahedral configuration with 15872 charges ($h=k=23$). The 
energy of
this configurations is \hbox{$E_i=-1.10562321$.}}
\label{fig3}
\end{figure}

\begin{figure}
\epsfxsize=.85\hsize
\begin{center}
\leavevmode
\epsfbox{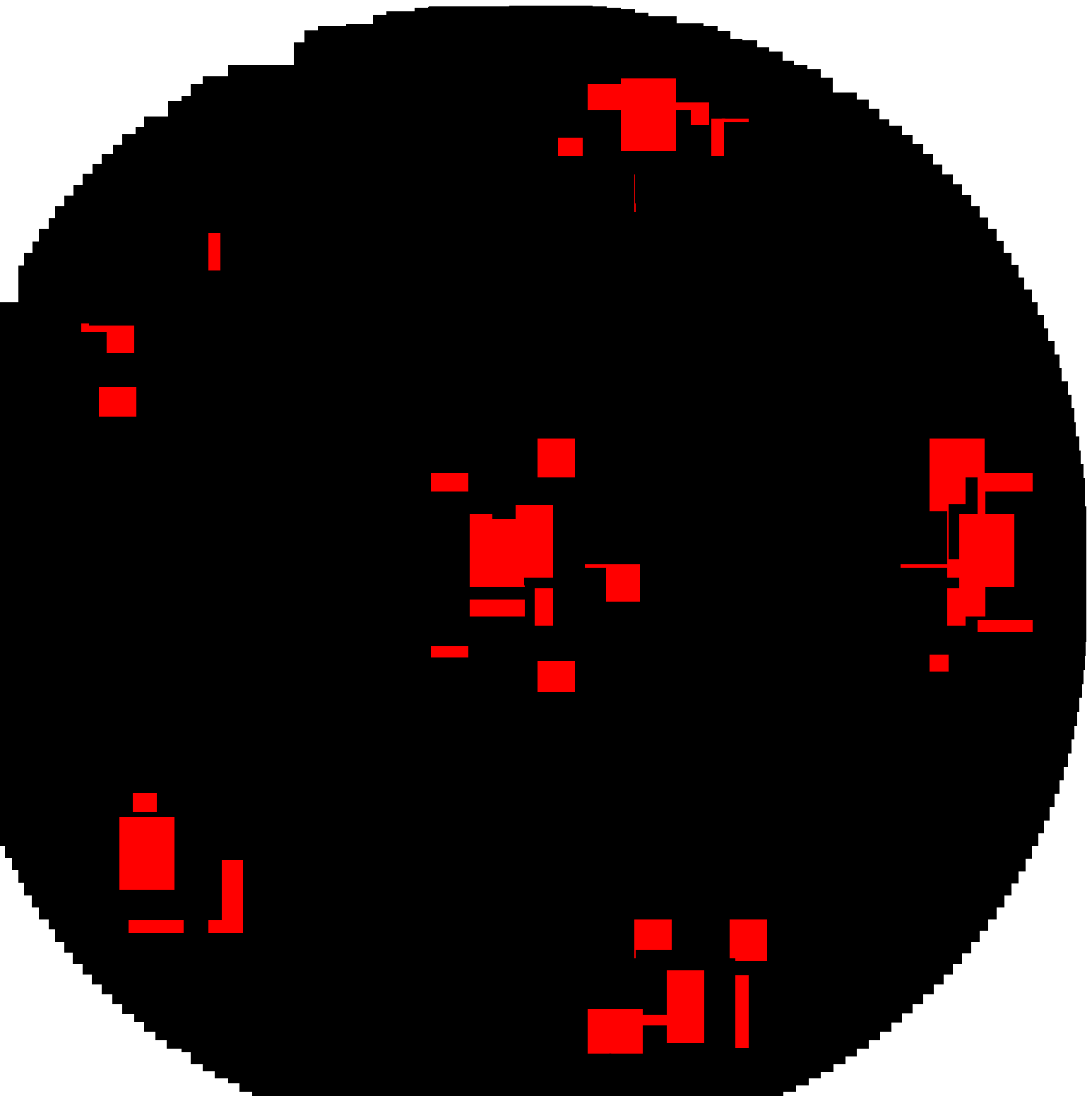}
\end{center}
\caption{15152 charges after relaxation. Largest spots correspond to 
seven-fold
disclinations, medium spots represent five-fold disclinations and 
smallest
spots are normal six-fold coordinated charges. This configuration is 
obtained
after removing the 3rd and 9th rings ($\equiv 720$ charges)  from  the
icosadeltahedral configuration with 15872 charges ($h=k=23$). The 
energy of
this configurations is \hbox{$E_i=-1.10561917$.}}
\label{fig4}
\end{figure}

We examine first configurations with $h=k$. For them the final 
configurations
obtained after energy relaxation following ring removal have 
dislocations on
the lines between the disclinations as envisaged in \cite{PD97}. Fig.\
\ref{fig2} is an example of a configuration with ($h=k=20$) minus the 
2nd and
9th rings around each disclination so containing 11342 charges. In 
Figs.\
\ref{fig3} and \ref{fig4} we have plotted  ($h=k=23$) minus 2nd and 
10th rings
in the former and minus 3rd and 9th rings in the latter. Dislocations 
obtained
by removing the second ring are bent forming pentagonal ``buttons"
\cite{TOOMRE} as in Figs.\ \ref{fig2} and \ref{fig3}. Fig.\ref{fig4} is 
very
similar to the kind of pattern which was suggested  in
\cite{PD97}.

\begin{figure}
\epsfxsize=.85\hsize
\begin{center}
\leavevmode
\epsfbox{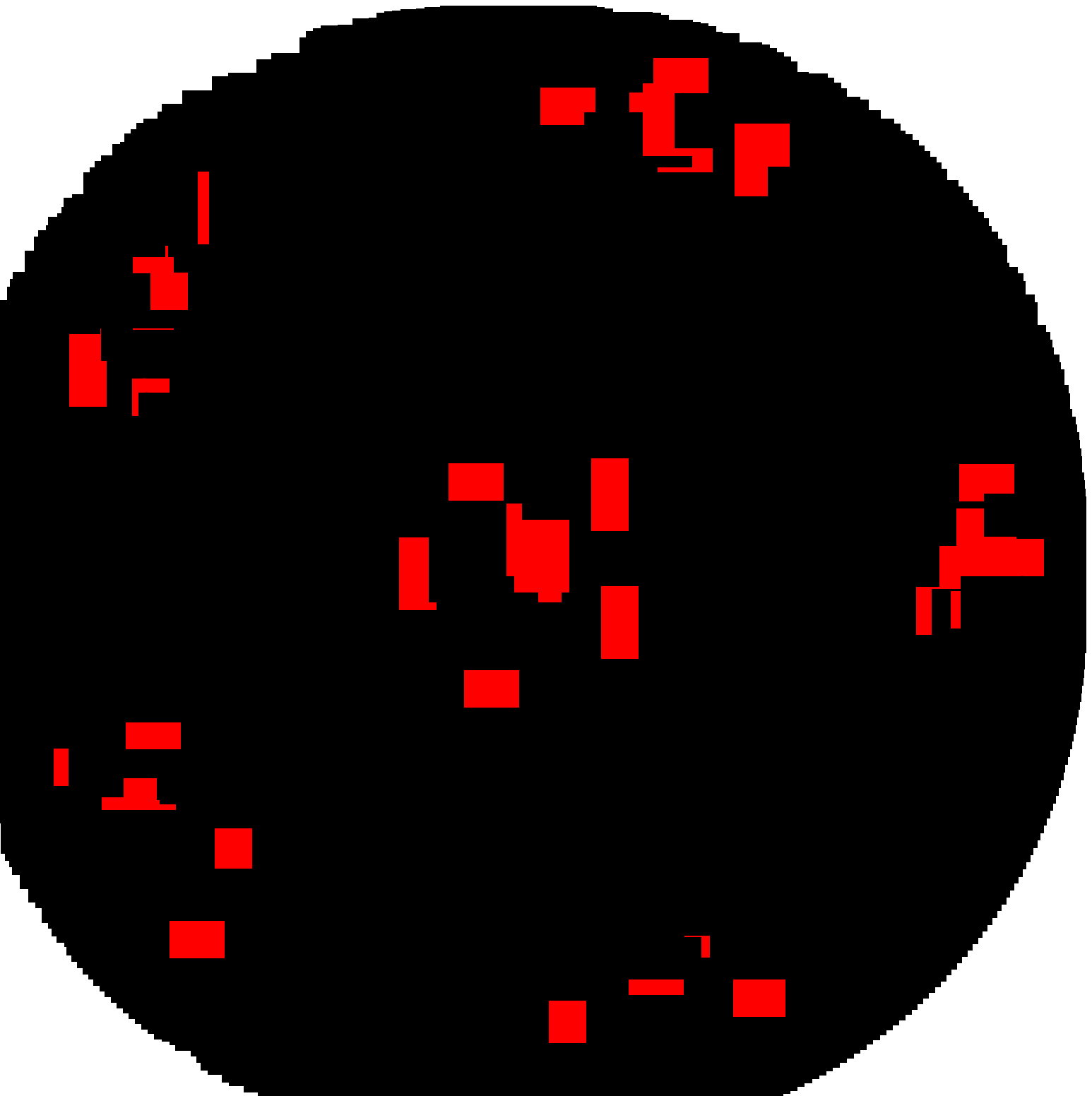}
\end{center}
\caption{15282 charges after relaxation. Largest spots correspond to 
seven-fold
disclinations, medium spots represent five-fold disclinations and 
smallest
spots are normal six-fold coordinated charges. This configuration is 
obtained
after removing the 2nd and 10th rings ($\equiv 720$ charges)  from the
icosadeltahedral configuration with 16002 charges ($h=40, k=0$). The 
energy of
this configurations is \hbox{$E_i=-1.10561047$.}}
\label{fig5}
\end{figure}

\begin{figure}
\epsfxsize=.85\hsize
\begin{center}
\leavevmode
\epsfbox{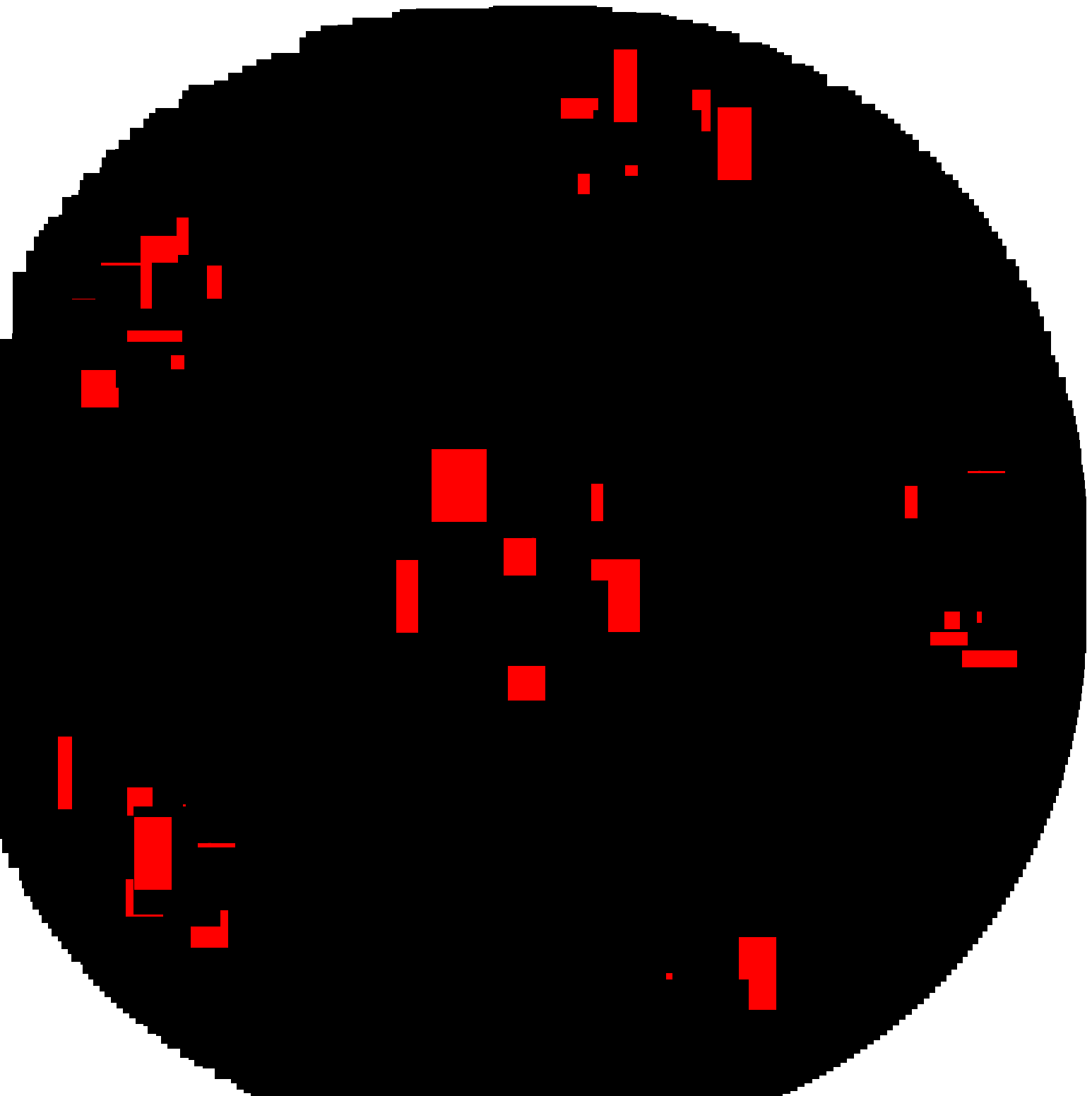}
\end{center}
\caption{15282 charges after relaxation. Largest spots correspond to 
seven-fold
disclinations, medium spots represent five-fold disclinations and 
smallest
spots are normal six-fold coordinated charges. This configuration is 
obtained
after removing the 3rd and 9th rings ($\equiv 720$ charges)  from the
icosadeltahedral configuration with 16002 charges ($h=40, k=0$). The 
energy of
this configurations is \hbox{$E_i=-1.10561147$.}}
\label{fig6}
\end{figure}

When $h\neq k$, the dislocations  obtained after removing rings are not 
on the
lines between disclinations but between these lines. The resulting 
patterns are
therefore of lower symmetry. In any of these three cases,
dislocations place themselves
onto a line rotated  an angle
$\theta$
from  the line between disclinations given by
\begin{equation}
\cos \theta = \frac{(h+k)(1+\cos 2\pi/5)}{\sqrt{(h^2+k^2+2hk\cos 
2\pi/5)
(2+2\cos 2\pi/5)
}}
\end{equation}
In Figs.\ \ref{fig5} and
\ref{fig6}  we study ($h=40,k=0$) minus 2nd and 10th rings in the 
former and
minus 3rd and 9th rings in the latter.

In conclusion, we have demonstrated that for certain values of $N$ 
there are
low energy arrangements of the charges which have full icosahedral 
symmetry.
For these special values of $N$ Thomson's problem seems to  reduce to 
the much
simpler task of finding which rings when removed minimize the energy.

APG  would like to acknowledge a grant and financial support from
CajaMurcia and EPSRC under grant GR/K53208. We thank A. Toomre
for telling us about his ``ring removal" method prior to its 
publication.


\begin{thebibliography}{99}

\bibitem{EH91} T.\ Erber and G.\ M.\ Hockney, J.\ Phys.\  A {\bf 24}
L1369 (1991).

\bibitem{AW94} E.\ L.\ Altschuler, T.\ J.\ Willians, E.\ R.\ Ratner, 
F.\

Dowla
 and
F.\ Wooten, Phys.\ Rev.\ Lett.\  {\bf 72} 2671 (1994).

\bibitem{PO96} A.\ P\'erez--Garrido, M.\ Ortu\~no, E.\ Cuevas and J.\
Ruiz,
J.\ Phys.\  A {\bf 29} 1973 (1996).

\bibitem{MD96} J.\ R.\ Morris, D.\ M.\ Deaven  and K.\ M.\ Ho,  Phys.\
Rev.\
 B {\bf 53} R1740 (1996).

\bibitem{EH97} T. Erber  and G. M. Hockney,
  Adv.\ Chem.\ Phys.\  {\bf 98}, 495 (1997).


\bibitem{PD97} A. P\'erez--Garrido, M. J. W. Dodgson and
M. A. Moore,  Phys.\ Rev.\ B {\bf 56}, 3640 (1997).

\bibitem{AW97} E. L. Altschuler, T. J. Williams, E. R. Ratner, R.
Tipton,
R. Stong, F. Dowla and F. Wooten,
 Phys.\ Rev.\  Lett.\ {\bf 78}, 2681 (1997).

\bibitem{PD97a}
A. P\'erez--Garrido, M. J. W. Dodgson,
M. A. Moore, M. Ortu\~no and A. D\'{\i}az--S\'anchez,
 Phys.\ Rev.\ Lett.\ {\bf 79}, 1417 (1997)


\bibitem{DM97} M.\ J.\ W.\ Dodgson and M.\ A.\ Moore,
 Phys.\ Rev.\  B {\bf 55}, 3816 (1997).

\bibitem{TOOMRE} A.\ Toomre, private communication and to be published.

\bibitem{BM77} L. Bonsall and A. A. Maradudin,
Phys.\ Rev.\ B {\bf 15}, 1959 (1977)

\end{thebibliography}
\end{document}